# Behavior of a Single-Atom Laser in the Sub-Poissonian Regime

Nikolay V. Larionov
*Institute of Machinery, Materials, and Transport; Institute of Physics, Nanotechnology and Telecommunications*
*Peter the Great St. Petersburg Polytechnic University*
Saint Petersburg, Russia
larionov.nickolay@gmail.com

*Abstract*—In the present work the model of a single-atom laser generating in the regime when incoherent pumping rate coincides with the cavity decay rate is theoretically investigated. Using the stationary equation for the phase averaged Glauber P function the analytical expressions for mean number of photons and its dispersion are obtained. In the limiting case of the strong-coupling regime the exact expression for the photon number distribution function is found. Obtained results describe the sub-Poissonian photon statistics and show an increase in mean number of photons to non-zero value when the pump tends to zero.

## I. INTRODUCTION

Nowadays, due to the development of quantum information technologies, quantum communications and measurements, systems consisting of microscopic or macroscopic quantum objects, interacting with the electromagnetic field, are of great interest [1-3]. In quantum optics, one of the most well-known microscopic systems is a single-atom laser. Its implementation can be represented by two-level atom including the pumping and interacting with a single damping cavity mode. Many theoretical papers have been devoted to this problem [3-8] (see references in these papers), in which various quantum effects inherent in this laser have been revealed: self-quenching for very strong external pumping; thresholdless generation and lasing without inversion for strong-coupling regime; vacuum-Rabi doublet in the spectrum; squeezing; entanglement between the emitter and the field; optical phase bistability; sub-Poissonian photon statistics and antibunching effect. This model was also realized experimentally [9].

Despite the simplicity of the model for a single-atom laser, exact solution have not yet been found. Only some solutions for various limiting cases was obtained (see [3], [5-7] and references in them). In this paper, we also consider one of the special cases: the bath responsible for the spontaneous decay of the atom is ignored; the rate of incoherent pumping of the atom coincides with the rate of decay of the cavity mode. The difference from other works related to this problem, is in the selected special case when the Fermi-like character in the photon distribution manifest itself most strongly. We also provide analytical expressions for the average number of photons and its dispersion and in the limiting case of the strong-coupling regime we obtain the exact solution for the photon distribution function.

## II. THEORETICAL MODEL OF A SINGLE-ATOM LASER

The laser model is represented by a single two-level atom interacting with a single damping cavity mode. The atom also interacts with two baths. The first bath causes spontaneous decay from the upper level $|2\rangle$ to the lower level $|1\rangle$ and the second one - the incoherent pumping from the lower level $|1\rangle$ to the upper level $|2\rangle$. The entire model is described by four constants: $g$ - atom-cavity mode coupling constant, $\kappa/2$ - cavity mode decay rate, $\gamma/2$ - rate of the atomic spontaneous emission out of the cavity mode, $\Gamma/2$ - rate of incoherent pumping.

The master equation for density operator $\hat{\rho}$ is

$$\frac{\partial \hat{\rho}}{\partial t} = -\frac{i}{\hbar}\left[\hat{V},\hat{\rho}\right] + \frac{\kappa}{2}\left(2\hat{a}\hat{\rho}\hat{a}^\dagger - \hat{a}^\dagger\hat{a}\hat{\rho} - \hat{\rho}\hat{a}^\dagger\hat{a}\right)$$
$$+ \frac{\gamma}{2}\left(2\hat{\sigma}\hat{\rho}\hat{\sigma}^\dagger - \hat{\sigma}^\dagger\hat{\sigma}\hat{\rho} - \hat{\rho}\hat{\sigma}^\dagger\hat{\sigma}\right) + \quad (1)$$
$$+ \frac{\Gamma}{2}\left(2\hat{\sigma}^\dagger\hat{\rho}\hat{\sigma} - \hat{\sigma}\hat{\sigma}^\dagger\hat{\rho} - \hat{\rho}\hat{\sigma}\hat{\sigma}^\dagger\right),$$

where $\hat{a}^\dagger, \hat{a}$ are the photon annihilation and creation operators in the cavity mode, $\hat{\sigma} = |1\rangle\langle 2|$ ($\hat{\sigma}^\dagger = |2\rangle\langle 1|$) is the operator of polarization of the two-level atom and $\hat{V} = i\hbar g\left(\hat{a}^\dagger\hat{\sigma} - \hat{\sigma}^\dagger\hat{a}\right)$ is the interaction operator between atom and cavity mode. Use (1) one can obtain the system of equations for Glauber $P$ function and additional quasiprobabilities. In the stationary case the following system of equations for the phase averaged quasiprobabilities can be derived [6,7]

$$\begin{cases} 2\tau IP - \left[(\omega-\eta)P - (\omega+\eta)D\right] = \tau\frac{\partial}{\partial I}I(P+D), \\ IP(\omega+\eta+\tau) - \frac{1}{\tau}\left[ID + \frac{1}{2}(P+D)\right] = \\ \frac{\partial}{\partial I}\left[2\tau I^2 P - \frac{1}{\tau}I(P+D)\right], \end{cases} \quad (2)$$

where we determined three dimensionless constants $\omega = \Gamma/2g$, $\eta = \gamma/2g$, $\tau = \kappa/2g$. Variable $I$ is the absolute square of complex number $z = \sqrt{I}\exp(i\varphi)$ which corresponds to the coherent state $|z\rangle$.

$P(I) = (1/2\pi)\int_0^{2\pi} P(I,\varphi)d\varphi$ is the phase averaged Glauber $P$ function and additional quasiprobability $D(I) = (1/2\pi)\int_0^{2\pi} D(I,\varphi)d\varphi$ relates to the mean value for the atomic inversion $\langle D \rangle$ as follows $\langle D \rangle = \pi \int_0^\infty D(I)dI$. We shall use (2) to get some important expressions.

At first let us obtain relation between mean number of photons in the cavity $\langle n \rangle = \pi \int_0^\infty IP(I)dI$ and the average of the square of the number of photons $\langle n^2 \rangle = \langle n \rangle + \pi \int_0^\infty I^2 P(I)dI$. On carrying out the integrations over $I$, (2) reduces to

$$\begin{cases} 2\tau\langle n\rangle - [(\omega-\eta)-(\omega+\eta)\langle D\rangle] = 0, \\ \langle n\rangle(\omega+\eta+\tau) - \frac{1}{\tau}\left[\langle ID\rangle + \frac{1}{2}(1+\langle D\rangle)\right] = 0, \end{cases} \quad (3)$$

where $\langle ID \rangle = \pi\int_0^\infty ID(I)dI$. To find $\langle ID \rangle$ we express $D(I)$ from the first equation of (2) as $D(I) = [(\omega-\eta)P - 2\tau IP + \tau\partial I(P+D)/\partial I]/(\omega+\eta)$ and substitute it in the integral $\langle ID \rangle$. The result is

$$\langle ID\rangle = \frac{1}{(\omega+\eta+\tau)}\left[(\omega-\eta+\tau)\langle n\rangle - 2\tau\langle n^2\rangle\right]. \quad (4)$$

After substitution of (4) into the second equation in (3) and resolving this system in terms of $\langle n \rangle$ and $\langle n^2 \rangle$ the following relation can be obtained

$$\begin{cases} \langle n^2\rangle + A(\omega,\eta,\tau)\langle n\rangle - B(\omega,\eta,\tau) = 0, \\ A(\omega,\eta,\tau) = \frac{\omega+\eta+\tau}{2(\omega+\eta)} - \left(\frac{\omega-\eta+\tau}{2\tau} - \frac{(\omega+\eta+\tau)^2}{2}\right), \\ B(\omega,\eta,\tau) = \frac{\omega}{2\tau}\frac{(\omega+\eta+\tau)}{(\omega+\eta)}. \end{cases} \quad (5)$$

Relation (5) was first obtained by Agarwal and Dutta Gupta [10], who used the method of continued fraction for this purpose.

Next expression we need is the second-order differential equation for $P(I)$ which one can easily derive from (2) [6,7]

$$\left(a_{02}I^2 + a_{03}I^3\right)\frac{\partial^2 P}{\partial I^2} + \left(a_{10} + a_{11}I + a_{12}I^2\right)\frac{\partial P}{\partial I} + \left(a_{20} + a_{21}I + a_{22}I^2\right)P = 0, \quad (6)$$

$a_{02} = \frac{\tau^3}{2}(\tau-\omega-\eta), \; a_{03} = \tau^4, a_{10} = \frac{\omega}{4}(\tau-\omega-\eta)$,

$a_{11} = \frac{\tau}{4}[3\eta^2\tau + 9\tau^3 + 4\omega - 12\tau^2\omega + \eta(2-12\tau^2+6\tau\omega) + \tau(3\omega^2-2)]$, $a_{12} = \frac{\tau^2}{2}(7\tau^2 - 3\tau\eta - 3\tau\omega - 2)$,

$a_{20} = \frac{1}{4}[6\tau^4 + \omega^2 - \eta^3\tau - 11\tau^3\omega - \tau\omega^3 + \eta^2(6\tau^2-3\tau\omega-1) + \eta\tau(12\tau\omega+4-11\tau^2-3\omega^2) + \tau^2(6\omega^2-3)]$, $a_{21} = \frac{\tau}{2}[\eta^2\tau + 3\tau^3 - 2\omega - 4\tau^2\omega + \tau\omega^2 + 2\eta\tau(\omega-2\tau)]$, $a_{22} = \tau^2$.

If we integrate (6) over $I$ and compare result with (5) we obtain the following boundary condition for the phase averaged $P$ function

$$P(0) = \frac{2\tau(\omega+\eta-\tau)}{\pi\omega(\omega+\eta)} \times \left[\langle n\rangle - \left(\frac{(\omega-\eta)}{2\tau} - \frac{(\omega+\eta)(\omega+\eta-\tau)}{2}\right)\right]. \quad (7)$$

This boundary condition can be used for investigation of (6).

The last expressions we need are expressions that bind $\langle n\rangle$, $\langle n^2\rangle$, $\langle n^3\rangle$ and $\langle n^4\rangle$. To obtain it the (6) should be multiply by $I^n$ ($n=1,2$) and then integrate over $I$. This manipulation gives two coupled relations

$$\begin{cases} a_{22}\langle n^3\rangle + (12a_{03} - 3a_{12} + a_{21} - 3a_{22})\langle n^2\rangle + (6a_{02} \\ -12a_{03} - 2a_{11} + 3a_{12} + a_{20} - a_{21} + 2a_{22})\langle n\rangle - a_{10} = 0, \\ a_{22}\langle n^4\rangle + (20a_{03} - 4a_{12} + a_{21} - 6a_{22})\langle n^3\rangle + (-60a_{03} \\ -3a_{11} + 12a_{12} + a_{20} - 3a_{21} + 12a_{02} + 11a_{22})\langle n^2\rangle + \\ +(40a_{03} + 3a_{11} - 8a_{12} - a_{20} + 2a_{21} - 12a_{02} - 2a_{10} \\ -6a_{22})\langle n\rangle = 0. \end{cases} \quad (8)$$

In the next section from (5) and (8) we get an analytical solution for $\langle n\rangle$ and $\langle n^2\rangle$. In the limit of strong-coupling regime (6) allows us to obtain exact expression for the photon number distribution function.

### III. RESULTS AND DISCUSSION

We are interested in the situation when $\eta = 0$ and when the cavity decay rate is equal to the rate of incoherent pumping of the atom, i.e. $\tau = \omega$. In this case the coefficients in (5), (6), (8) are

$$A(\tau,0,\tau) = 2\tau^2, B(\tau,0,\tau) = 1;$$
$$a_{02} = 0,\; a_{03} = \tau^4, a_{10} = 0,\; a_{11} = \tau^2/2, \quad (9)$$
$$a_{12} = \tau^2(2\tau^2-1), a_{20} = -\tau^2/2,\; a_{21} = -\tau^2,\; a_{22} = \tau^2.$$

And from (3) and (5) we have the following conditions

$$\langle n\rangle = \frac{1-\langle D\rangle}{2} \le 1,\; \langle n\rangle \le 1/2\tau^2. \quad (10)$$

Inequalities (10) allow us to neglect some small terms in (8). Let us call the first order approximation if we neglect all terms which are proportional to $\langle \hat{a}^\dagger \hat{a}^\dagger \hat{a}^\dagger \hat{a}\hat{a}\hat{a} \rangle$. And the second order approximation will be correspond to the neglecting of terms arising from $\langle \hat{a}^\dagger \hat{a}^\dagger \hat{a}^\dagger \hat{a}^\dagger \hat{a}\hat{a}\hat{a}\hat{a} \rangle$. Thus, in the second order approximation to find analytical expressions for $\langle n \rangle$ and $\langle n^2 \rangle$ we should to resolve the following system of equations, obtained from (5) and (8)

$$\begin{cases} \langle n^2 \rangle + 2\tau^2 \langle n \rangle - 1 = 0, \\ \langle n^3 \rangle + (6\tau^2 - 1)\langle n^2 \rangle - \left(6\tau^2 + \frac{3}{2}\right)\langle n \rangle = 0, \\ (12\tau^2 + 3)\langle n^3 \rangle - (36\tau^2 + 11)\langle n^2 \rangle + \\ + (24\tau^2 + 8)\langle n \rangle = 0. \end{cases} \quad (11)$$

Solution of (11) is

$$\begin{cases} \langle n \rangle = \dfrac{4(4 + 21\tau^2 + 36\tau^4)}{25 + 8\tau^2(19 + 39\tau^2 + 36\tau^4)}, \\ \langle n^2 \rangle = \dfrac{(5 + 12\tau^2)^2}{25 + 8\tau^2(19 + 39\tau^2 + 36\tau^4)}. \end{cases} \quad (12)$$

In the Fig. 1 we present $\langle n \rangle$ and the Mandel $Q$-parameter $Q = (\langle n^2 \rangle - \langle n \rangle^2)/\langle n \rangle - 1$ calculated using (12) and numerical simulations of (1). One can see that in the second order approximation the curves are very close to numerical results. The best sub-Poissonian statistics takes place for $\tau \approx 1/\sqrt{2}$ and $Q = -0.15$. This result is in agreement with results from [8] and can be interpreted as Fermi-like character in the photon distribution for considered case: during photon lifetime in the cavity it manages to interact with an atom once $\kappa \approx \sqrt{2}g$. For $\tau \gg 1$ ($\kappa \gg g$ - weak coupling regime) the interaction between an atom and photon during its lifetime $\kappa^{-1}$ has the rare character. At the same time the pumping rate also increases and this leads to self-quenching effect when the atom is not able to emit. Indeed, atom trapping on the upper level following from (10): $\langle n \rangle = (1 - \langle D \rangle)/2$. Another specific situation occurs when $\tau = \omega \to 0$: the mean number of photons and $Q$-parameter do not tend to zero. This situation corresponds to the strong coupling regime $g \gg \kappa = \Gamma$: the photon is very strong coupling with an atom and it is hard for it to leave the cavity. Simultaneously with the photon lifetime the atom lifetime in the ground state $\Gamma^{-1}$ is increased to infinity while $\tau = \omega \to 0$. All transitions from the lower atomic level to the upper and inverse are primarily caused by coherent interaction with intracavity long-lived photon. In this case, populations of atomic levels become almost equal to each other and situation begins to resemble non-damped Rabi oscillation problem.

It is interesting, that in this limiting case $\tau = \omega \to 0$ the equation for the $P$ function can be easily solved. Indeed, from (6) when $\eta = 0, \tau = \omega \to 0$ we obtain

$$\begin{cases} \left(\dfrac{1}{2}I - I^2\right)\dfrac{\partial P}{\partial I} + \left(-\dfrac{1}{2} - I + I^2\right)P = 0, \\ P(I) = C_0 \dfrac{I \cdot \exp(I)}{(1 - 2I)^{3/2}}. \end{cases} \quad (13)$$

Solution (13) can be derived from the solution found in [5] (or in [6]), where authors discussed its properties. Because of difficulties to work with quasi-probability distribution let us transform (13) into equation for the photon distribution function $\rho(n) = (1/n!)\int_0^\infty P(I)\exp(-I)I^n dI$. After simple math from the equation for the $P$ function we obtain

$$\rho(n+1)\left\{(n+1)(n+2) - \frac{1}{2}(n+1)\right\} \\ -\rho(n)\frac{(n+1)}{2} = 0. \quad (14)$$

The normalized exact solution of (14) is

$$\rho(n) = C \frac{2^{-n}(n+1)}{\Gamma\left(\frac{3}{2} + n\right)}, \quad C = \frac{\sqrt{\pi}}{1 + \sqrt{2\pi e}\,\text{erf}\left(\frac{1}{\sqrt{2}}\right)}. \quad (15)$$

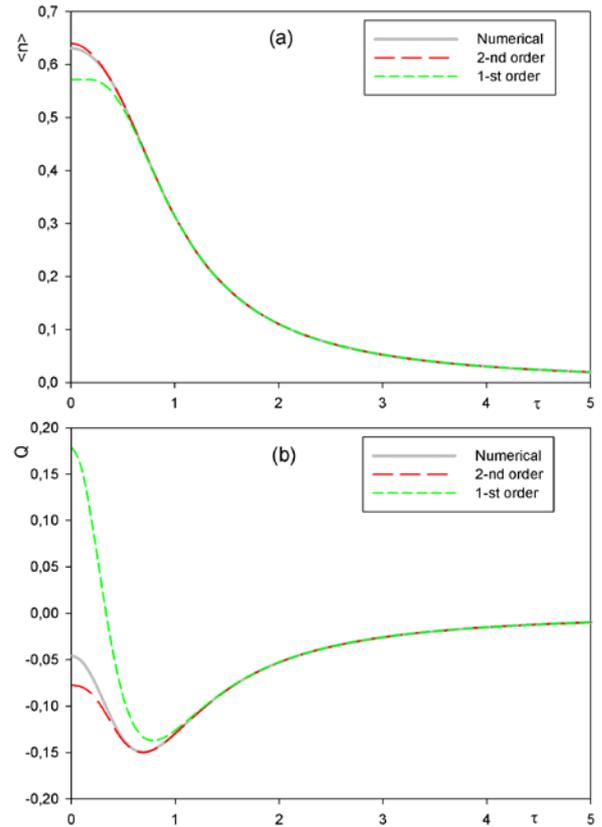

Fig. 1. The mean number of photons in the cavity mode (a) and corresponding Mandel $Q$-parameter (b) vs $\tau$. Solid gray line - numerical calculation of (1); Red long-dash line – analytical results obtained in the 2nd order approximation (12); Green short-dash line – analytical results obtained in the 1st order approximation.

Now it is easy to find normalization constant $C_0$ for quasi-probability (13). Using well known connection between $\hat{\rho}$ and $P$ -

$P(z,z^*) = e^{|z|^2}\int \langle -\beta|\hat{\rho}|\beta\rangle e^{|\beta|^2} e^{-\beta z^* + \beta^* z} d^2\beta/\pi^2$, we obtain $C_0 = -2\left[\pi + \pi\sqrt{2\pi e}\,\text{erf}\left(1/\sqrt{2}\right)\right]^{-1}$, where $\text{erf}(x)$ is the error function. This result is in agreement with [5].

Using (15) we can compare our result (12) for $\tau = 0$ and numerical simulations with exact one: $\langle n\rangle = \sum_{n=0}^{\infty} n\rho(n) = 0.630843$, $\langle n^2\rangle = \sum_{n=0}^{\infty} n^2\rho(n) = 1$, $Q = -0.0456627$. From (12): $\langle n\rangle = 0.64$, $\langle n^2\rangle = 1$, Q=-0.0775. It should be noted, that such a stationary state (15) is probably possible, but the time to reach it tends to infinity when $\tau = \omega \to 0$. This aspect manifests itself in numerical simulation of a dynamic problem.

In the Fig. 2 we plot (15).

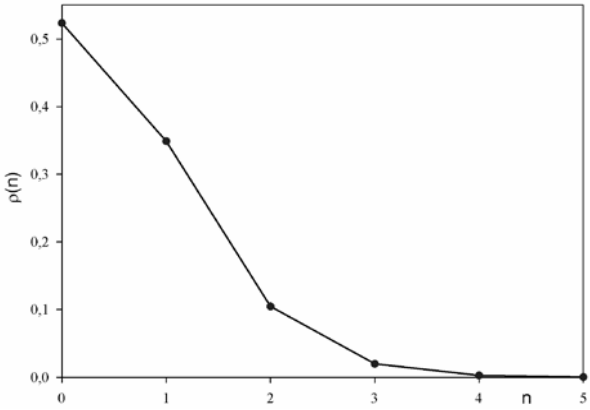

Fig. 2. The photon distribution function (15).

## IV. Conclusion

In this paper, we investigated a simple model of a single-atom laser that generated in the regime of equality of the atomic pumping rate and the decay rate of the cavity, $\Gamma = \kappa$. Spontaneous decay of the higher atomic level was ignored. Based on the stationary equations for the phase averaged quasi-probabilities (2), (6) the system of equations for different moments of the field operators was obtained. This system allowed us to derive analytical solution for mean number of photons and its dispersion as functions of the $\tau = \kappa/2g$ (12). These analytical results describe well the sub-Poissonian behavior of a single-atom laser, which is confirmed by comparison with numerical simulations of the dynamic problem (1).

The transition to the regime of strong coupling $\tau \to 0$ is accompanied by increasing of the mean number of photons in the cavity, which tends to nonzero value $\langle n\rangle \approx 0.63$. For this limiting case, an exact solution for the photon distribution function $\rho(n)$ (15) was found. This solution also confirms our results.

### References

[1] A. S. Kuraptsev, I. M. Sokolov, K. A. Barantsev, A. N. Litvinov, E. N. Popov, "Interatomic Dipole–Dipole Interaction in a Fabry–Perot Cavity with Charged Mirrors", Bull. Russ. Acad. Sci. Phys., vol. 83, pp. 242-246, 2019.

[2] A. S. Kuraptsev, I. M. Sokolov, "Light trapping in an ensemble of pointlike impurity centers in a Fabry-Perot cavity", Phys. Rev. A, vol. 94, 022511, 2016.

[3] N. V. Larionov, M. I. Kolobov, "Analytical results from the quantum theory of a single-emitter nanolaser", Phys. Rev. A, vol. 84, 055801, 2011.

[4] Y. Mu, C. M. Savage, "One-atom lasers", Phys. Rev. A, vol. 46, 5944, 1992.

[5] S. Y. Kilin, T. B. Karlovich, "Single-atom laser: Coherent and nonclassical effects in the regime of a strong atom-field correlation", J. Exp. Theor. Phys., vol. 95, pp. 805-819, 2002.

[6] N. V. Larionov, M. I. Kolobov, "Quantum theory of a single-emitter nanolaser," Phys. Rev. A, vol. 88, 013843, 2013.

[7] E. N. Popov, N. V. Larionov, "Glauber-Sudarshan P function in the model of a single-emitter laser generating in strong coupling regime", Proceeding of SPIE, 9917, 99172X, 2016.

[8] A. V. Kozlovskii, A. N. Oraevskii, "Sub-Poissonian radiation of a one-atom two-level laser with incoherent pumping" J. Exp. Theor. Phys., vol. 88, pp. 666–671, 1999.

[9] J. McKeever, A. Boca, A. D. Boozer, J. R. Buck, H. J. Kimble, "Experimental realization of a one-atom laser in the regime of strong coupling", Nature, vol. 425, 268, 2003.

[10] G. S. Agarwal, S. Dutta Gupta, "Steady states in cavity QED due to incoherent pumping", Phys. Rev.A, vol. 42, 1737, 1990.